\documentclass[superscriptaddress,twocolumn]{revtex4}
\usepackage{graphicx}
\usepackage{color}
\usepackage{amsmath}
\usepackage{multirow}
\usepackage{booktabs}
\usepackage{ulem}
\newcommand{\be}{\begin{equation}}
\newcommand{\ee}{\end{equation}}
\newcommand{\bea}{\begin{eqnarray}}
\newcommand{\eea}{\end{eqnarray}}

\begin{document}

\title{New algorithm for Multi-Reference Unitary Coupled Cluster quantum computations of molecule systems}

\author{Di Wu} 
\affiliation{College of Physics and Electronic Information Engineering, Neijiang Normal University, Neijiang 641100, China
} 

\author{C.L. Bai}
\affiliation{College of Physics, Sichuan University, Chengdu 610065, China
} 

\author{H. Sagawa}
\affiliation{
RIKEN Nishina Center, Wako 351-0198, Japan}
\affiliation{
Center for Mathematics and Physics,  University of Aizu, 
Aizu-Wakamatsu, Fukushima 965-8560,  Japan}

\author{H.Q. Zhang}
\affiliation{China Institute of Atomic Energy, Beijing 102413, China
}

\begin{abstract}
We present a new efficient algorithm for Multi-Reference Unitary Coupled Cluster (MR-UCC) approach
that integrates quantum computing techniques with 
particle number conserved (PNC) gates.  
This algorithm  makes possible to use  sophisticated MR wavefunction without any heavy burden on the 
computer resources to calculate the ground state energies of LiH, BeH$_2$, and H$_6$.   The higher order 
correlations in the subsequent UCC calculations can be included taking only Singles (S) and Doubles (D) multiplets 
of the wave operator.
This algorithm achieves high accuracy results with minimal resources compared with other available sophisticated 
quantum computational models, using only a single quantum circuit across all bond length, including dissociation. 
\end{abstract}

\maketitle

\section{introduction}\label{intro}
The ground state energies of molecules are the key ingredients to the study of 
chemical systems and their reactions, including photochemical reaction and 
ultracold molecules \cite{warshel,karman}. 
Its accurate and efficient calculations
are at the forefront of quantum computing applications in the Noisy Intermediate-Scale Quantum (NISQ) era characterized by quantum processors containing up to 1,000 qubits
\cite{preskill,cerezo}.
The Variational Quantum Eigensolver (VQE) algorithms have been adopted as a leading approach in this domain \cite{peruzzo,mcclean,li19}.
These algorithms primarily focus on calculating the important electron correlations 
to achieve the chemical accuracy \cite{fedorov,tilly,blekos,fauseweh,grimsley,ryabinkin,ryabinkin20,zhao23,shang,sun23,tazi}.
However, 
the VQE algorithm is suffered by escalating  demand on computational resource to aquire the numerical 
precision as the model scale increases \cite{grimsley}.
Moreover, as the electron correlations vary with bond length, the existing VQE algorithms 
 require often
bespoken quantum circuits for different bond lengths, 
particularly in dissociation regions.
From the viewpoint of the configuration interaction (CI) and 
coupled cluster theories (CC), one need to
include the high-order multiplets to solve this problem, such as the triples (T)
 and even quadruples (Q) 
multiplets, which lead to a significant increase in computing costs \cite{noga,scuseria,raghavachari}.

Alternatively, the multi-reference (MR) methods  
offer a platform to account for higher-order correlations, which make possible the model  to 
enhance the numerical accuracy
 with only lower-order multiplets of the wave operator \cite{lyakh,oliphant,yao}.
In principle,  the more comprehensive correlations were considered in the MR wavefunction,  
the lower multiplets 
of the wave operator were  required in the subsequent energy evaluation. However, this approach demands either
a great number of determinants within the MR wavefunction, or prior 
knowledge of the system to select appropriate fewer determinants, which significantly increase the complexity of the
 calculations.  In reality, one uses simple MR wavefunction, which is in general not enough to include  higher-order multiplets of wave operator to attain 
the desired precision \cite{lischka,moraes,lee,sugisaki,greene,otten,kottmann,evangelista,evangelista2010,hanauer,sun24x}.

In this work, 
we present a new quantum computing technique of the MR-UCC calculation with the particle number conserved (PNC) 
circuit \cite{anselmetti},  
which enables to use the sophisticated MR wavefunction with a large number of determinants without any heavy 
 resource requirement.
In our algorithm,   the PNC
circuit generates automatically the MR wavefunction, which attains roughly the chemical accuracy.  This is the key issue of our new algorithm.
As enough electron correlations are then included, the subsequent UCC calculation, 
utilizing only S and D multiplets,
can attain high accuracy across the whole bond length with an uniform quantum circuit. 

This paper is organized as follows: In Sec. \ref{meth} we introduce the framework of MR-UCC method. In Sec. \ref{res_disc}, the calculating results of 
LiH, BeH$_2$, and H$_6$ are shown. Sec. \ref{sum} is devoted to the summary.

\section{methods}\label{meth}
We start from the Hamiltonian of the molecule in term of the second quantization representation, 
\begin{equation}\label{H}
\hat{\mathcal{H}}=\sum_{ik}h_{ik}a_i^\dagger a_k+\frac{1}{2}\sum_{ijkl}h_{ijkl}a_i^\dagger a_j^\dagger a_la_k,
\end{equation}
where $h_{ik}$ and $h_{ijkl}$ are one- and two- electron integrals, 
respectively. 
In the quantum processors, 
$a^\dagger,a$ are transformed into Pauli strings via the 
Jordan-Wigner transformations \cite{jordan}. 
Thus $\hat{\mathcal{H}}$ is rewritten in the following form,
\begin{equation} 
\hat{\mathcal{H}}=\sum_{i,k}c_k^i\sigma_k^i+\sum_{ijkl}c_{kl}^{ij}\sigma_k^i\otimes\sigma_l^j+\cdots. \label{Hpa}  
\end{equation}
The $\sigma_k^i$ in Eq. (\ref{Hpa}) stands for the Pauli operators or identity, $\sigma_k^i=(\sigma_x,\sigma_y,\sigma_z, I)$,  that acts on the $i$th qubit, and  
$c$ are the corresponding transformation coefficients from Eq. \eqref{H} to Eq. \eqref{Hpa}. 

\begin{figure}[htbp]
\includegraphics[scale=0.26]{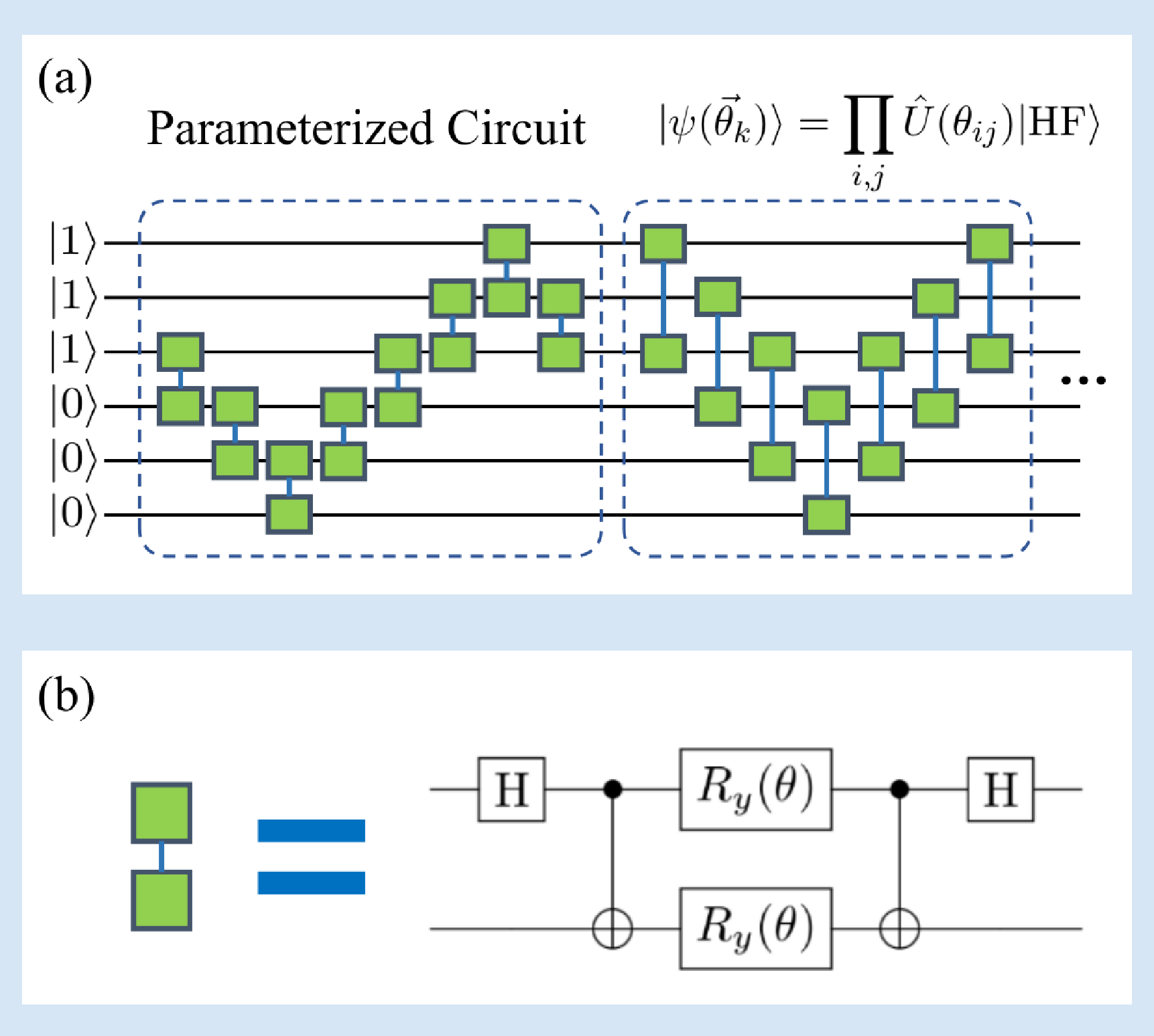}
\caption{(a) The structure of quantum circuit to generate MR wavefunction, where $\hat{U}(\theta_{ij})$ is represented 
with the green square. 
(b) The details of particle number conserved circuit $\hat{U}(\theta)$ \cite{anselmetti}, which
consists of two Hadamard gates (H), two CNOT gates, and two rotating Y gates ($R_y(\theta)$)
with angle $\theta$, respectively.} \label{workflow}
\end{figure}

The MR-UCC model can be divided into two stages as follows.
The first stage is to construct the MR wavefunction $|\psi(\vec{\theta})\rangle$ within the VQE framework. 
Starting from the Hartree-Fock
(HF) state, $|\psi(\vec{\theta})\rangle$ can be formulated mathematically as 
\begin{equation}
|\psi(\vec{\theta})\rangle=\prod\limits_{i,j}\hat{U}(\theta_{ij})|\mbox{HF}\rangle, \label{MRS}
\end{equation}
where $i,j$ denote the $i$th and $j$th qubits.  The operator  $\hat{U}(\theta)$ (see Fig. \ref{workflow}(b)) is
a parameterized PNC circuit\cite{anselmetti}, 
\begin{align}
&\hat{U}(\theta)|01\rangle=\cos\theta|01\rangle+\sin\theta|10\rangle\label{u01},
\end{align}
where $\theta$ is the circuit parameter,
1 and 0 denote either occupied or unocuupied single-particle states in HF approximation. 
In Eq. \eqref{u01} the one-particle one-hole (1p-1h) configuration is  generated.  
By applying this circuit on different 
pair of qubits, a series of $n$p-$n$h configurations are created.

The architecture of VQE circuit employed for constructing $|\psi(\vec{\theta})\rangle$ 
is depicted in Fig. \ref{workflow}(a). 
We first act the PNC circuit on the adjacent qubits, then proceed to the next-nearest-neighbor 
qubits continouesly 
to achieve the acceptable accuracy. 
The circuit parameters are optimized by minimizing the expectation value of the Hamiltonian $\langle \hat{\mathcal{H}}\rangle$ with the Adam algorithm. 

The second stage involves the application of the wave operator $e^{\hat{A}(\vec{c})}$ on the MR wavefunction $|\psi(\vec{\theta})\rangle$,
\be
|\Psi(\vec{\theta}, \vec{c})\rangle=e^{\hat{A}(\vec{c})}  |\psi(\vec{\theta})\rangle.
\ee
In this work, we take the same form of $e^{\hat{A}(\vec{c})}$ as that in UCCSD \cite{ucc} in which the wave operator is expressed by the cluster operator as $\hat{A}(\vec{c})=\hat{T}(\vec{c})-\hat{T^\dagger}(\vec{c})$,  and 
the cluster operator $\hat{T}(\vec{c})$ 
takes into account S and D multiplets,
\begin{equation}
\hat{T}(\vec{c})=\sum_{mi}c_{mi}a^\dagger_ma_i+\sum_{mnij}c_{mnij}a^\dagger_ma^\dagger_na_ja_i,\label{eqT}
\end{equation}
where $m,n$ stand for the particle states, and $i,j$ are the hole states of the HF state.
The expectation value of the Hamiltonian operator in the 
presence of the UCCSD operator is,
\begin{align}
E(\vec{\theta},\vec{c})=&\langle \Psi(\vec{\theta}, \vec{c})|\hat{\mathcal{H}}|\Psi(\vec{\theta},\vec{c})\rangle
=\langle\psi(\vec{\theta})|e^{-\hat{A}(\vec{c})}\hat{\mathcal{H}}e^{\hat{A}(\vec{c})}|\psi(\vec{\theta})\rangle\nonumber \\
=&\langle\psi(\vec{\theta})|\hat{\mathcal{H}}'(\vec{c})|\psi(\vec{\theta})\rangle.\label{EE}
\end{align}
By utilizing the Baker-Campbell-Hausdorff (BCH) formula,  $\hat{\mathcal{H}}'(\vec{c})$ can be expanded to a first-order form
\begin{equation}\hat{\mathcal{H}}'(\vec{c})\approx\hat{\mathcal{H}}-[\hat{A}(\vec{c}),\hat{\mathcal{H}}].\label{bcheq}\end{equation}
Similar to Eq.  \eqref{Hpa}, $\hat{\mathcal{H}}'(\vec{c})$ is transformed to the form of Pauli strings 
that can be directly implemented on quantum computers. 
This transformation not only simplfies the energy evaluation effectively in the MR-UCCSD approach, but also 
reduces the complexity of the evaluation of $\hat{\mathcal{H}}'(\vec{c})$ by using the algebraic operations of Pauli strings.

According to Eq.  \eqref{EE}, $E(\vec{\theta},\vec{c})$ is evaluated using the wavefunction $|\psi(\vec{\theta})\rangle$
obtained in the first stage, which avoids the calling for extra CNOT gates, saving quantum computing resources.
The coefficients $c_{mi}$ and $c_{mnij}$ are optimized using the Adam algorithm.

\section{results and discussions}\label{res_disc}
The ground state energies of LiH, BeH$_2$, and H$_6$ are
calculated in order to test the performances of the MR-
UCCSD algorithm, which serves as a common benchmark
in the VQE calculations.
In the calculations, the electron integrals 
are calculated with  the Slater-type orbitals represented by three Gaussian functions (STO-3G) basis using the 
Python-based Simulations of Chemistry Framework (PySCF) \cite{pyscf}.
The numbers of qubit required for molcules LiH, BeH$_2$, and H$_6$ are 12, 14, and 12, respectively.

\begin{figure} [t]
\centering
\includegraphics[scale=1]{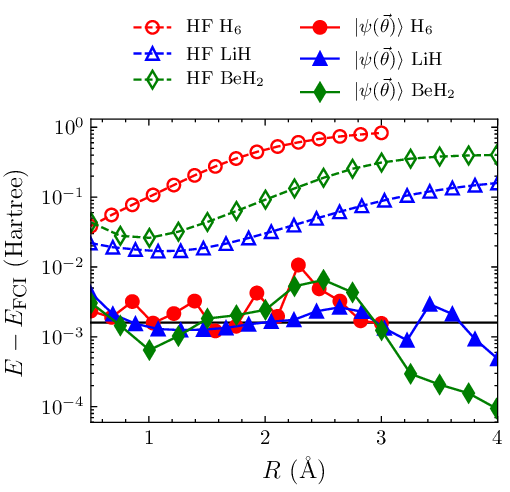}
\caption{The energy differences $E-E_{\rm FCI}$ for LiH, H$_6$, and BeH$_2$ between HF or the multi-reference states generated 
with PNC circuit and 
the FCI results as a function of the bond length $R$. The chemical accuracy is indicated by 
a black horizontal line in the figure.  
 See text for more details.}\label{error_mr}
\end{figure}

The computing errors of $\langle \hat{\mathcal{H}}\rangle$ calculated with the MR wavefunction
 $|\psi(\vec{\theta})\rangle$ and HF state (the reference state of the single-reference UCCSD) respect to the full configuration interaction 
(FCI) states (taken as the exact solution) are depicted in Fig. \ref{error_mr}. 
The computational precision of 
the multi-reference states 
is predominantly concentrated around the chemical accuracy along the whole bond length, 
2 to 3 orders of magnitude better than that of the HF states, particularly in the regions close to the dissociation. 

\begin{figure}[t]
\centering
\includegraphics[scale=1]{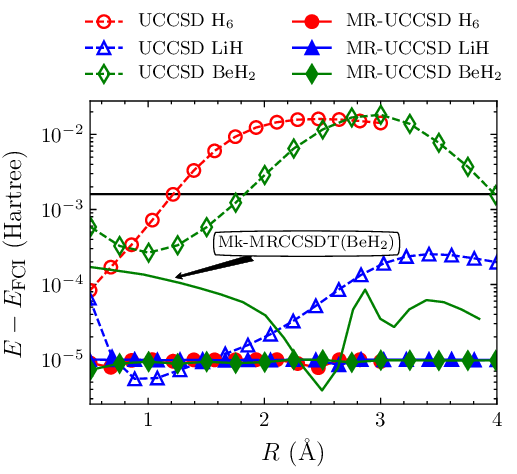}
\caption{The energy differences $E-E_{\rm FCI}$  between 
 UCCSD, MR-UCCSD, and Mk-MRCCSDT calculations,   and 
FCI results for LiH, H$_6$, and BeH$_2$ as a function of the bond length $R$.  The Mk-MRCCSDT's results are taken  
from Ref. \cite{evangelista2010}. The chemical accuracy is indicated by 
a black horizontal line in the figure. 
See text for more details.}\label{mrcc}
\end{figure}

\begin{table} [b]
\caption{Number of determinants and independent parameters for the present MR and the FCI wavefunctions.
 See text for more details.}\label{mr}
\begin{tabular}{l ccccc}
\hline
molecules & \multicolumn{2}{c}{No. det.} & & \multicolumn{2}{c}{No. Indep. Params.}\\
\cline{2-3} \cline{5-6}
& MR & FCI & & MR & FCI\\
\hline
LiH & 258 &  495  & & 54 & 495 \\
H$_6$ & 924 & 924 & & 260 & 924 \\
BeH$_2$ & 2174 & 3003 & & 198 & 3003\\
\hline
\end{tabular}
\end{table}

Table \ref{mr} lists the number of determinant utilized in the present MR methodology and the FCI, as well as the 
numbers of the circuit parameters employed to generate the MR wavefunction in Eq. \eqref{MRS} 
and the independent parameters in FCI wavefunction. 
Our approach includes from hundreds to thousands of 
determinants, which are the same for H$_6$ and  comparable for LiH and BeH$_2$ to the numbers of determinants  adopted in the exact FCI calculation.
This enriched MR wavefunction is able to include enough 
electron correlations in the wave function, and allows for the inclusion of only lower-order of multiplets, $S$ and $D$
in the subsequent UCC calculation, 
 without losing good high accuracy. However,  the number of the circuit parameters in the present  approach is much 
smaller than the numbers of determinants, which can be implemented by using a compact circuit,  and consequently saving the computing resources.

Fig. \ref{mrcc} presents the  ground state energy differences $E-E_{\rm FCI}$ for LiH, H$_6$, and BeH$_2$ between 
 MR-UCCSD, UCCSD, and Mk-MRCCSDT results \cite{evangelista2010}  and the FCI benchmark calculations.  The chemical accuracy  is indicated by 
a horizontal line in the figure. 
It is observed that near the equilibrium bond lengths (1.59 \AA~for LiH, 0.86 \AA~for H$_6$, 
and 1.25 \AA~for BeH$_2$), the single-reference UCCSD achieves chemical accuracy, with a maximum precision 
of about 10$^{-5}$ Hartree around the 1.0 \AA~bond length for  LiH.
However, as bond lengths increase, particularly approaching dissociation, the single-reference UCCSD
 fails to maintain chemical accuracy
for BeH$_2$ and H$_{6}$.   The numerical error increases also for LiH molecule although it is still below the chemical accuracy. 
These large errors can be attributed  to the strong correlations between multiple 
electron pairs in BeH$_2$ and H$_6$ 
 during the bond-breaking process \cite{grimsley}, 
where the single-reference UCCSD framework is inadequate to describe the molecules accurately. 
The MRCC approach suggested by Mukherjee and co-workers (Mk-MRCC) \cite{mk} has been viewed as one of the best 
algorithm among MR methods on classical computers for its size extensively.  
The Mk-MRCCSDT results for BeH$_2$\cite{evangelista2010} are also shown in Fig. \ref{mrcc}  by  a green solid line,
which utilize four determinants in the complete active space for the MR wavefunction, and the triple excitation
mode is also included in the cluster operator. Consequently, the precision is around 10$^{-4}$ Hartree, which is an
order of magnitude higher than the chemical accuracy across the entire bond length.
Since the highest precision in these methods is 10$^{-5}$ Hartree, we stop the iterations when the error becomes smaller than 
10$^{-5}$ Hartree in the MR-UCCSD calculations.   
Much  better accuracy of the present MR-UCCSD results 
especially near  the dissociation bond length 
 is attributed to the utilization of the  sophisticated MR wavefunctions, but standard UCCSD models with single reference states fails to get good accuracy .

\begin{table}[t]
\caption{The number of CNOT gates needed to achieve  the given accuracy (10$^{-5}$ Hartree) in various quantum computational models.
 See text for details.}\label{ncnot}
\begin{tabular}{lcccc}
\hline
Ansatz & Ref. & LiH & H$_6$ & BeH$_2$\\
\hline
pp-tUPS & \cite{burton} & 210 & 735 & $-$\\
\hline
QEB-ADAPT-VQE & \cite{burton} & $\sim$270 & $\sim$2000 & $-$\\
~ & \cite{yordanov} & $\sim$260 & $\sim$2250 & $\sim$880\\
~ & \cite{ramoa} & $\sim$280 & $\sim$2100 & $\sim$750\\
\hline
sQEB-ADAPT-VQE & \cite{sun24} & 50$\sim$200 & $\sim$1300 & $\sim$600\\
\hline
FEB-ADAPT-VQE & \cite{burton} & $\sim$400 & $\sim$2800 & $-$\\
~ & \cite{sun24} & $\sim$400 & $\sim$2500 & $\sim$1000\\
\hline
qubit-ADAPT-VQE & \cite{yordanov} & $\sim$320 & $\sim$2600 & $\sim$970\\
~ & \cite{ramoa} & $\sim$320 & $\sim$2400 & $\sim$1100\\
\hline
fermionic-ADAPT-VQE & \cite{yordanov} & $\sim$430 & $\sim$3300 & $\sim$920\\
\hline
QEB Gradient ADAPT & \cite{long} & $\sim$250 & $\sim$1850 & $\sim$750\\
\hline
CEO-ADAPT-VQE & \cite{ramoa} & $\sim$180 & $\sim$1000 & $\sim$500\\
\hline
MR-UCCSD & this work & 108 & 520 & 396\\
\hline
\end{tabular}
\end{table}

The number of CNOT gates utilized in our calculations and other VQE models to attain the accuracy of more than 10$^{-5}$ Hartree in
molecular energy are listed in Table \ref{ncnot}.
In the NISQ era, 
the number of CNOT gate is limited by the coherence time of entangled qubits.
There are two types of models which are extensively used in NISQ era: the pp-tUPS model, 
which integrates the perfect-pairing (pp) valence bond theory with the tiled  unitary 
product state (tUPS) approach \cite{burton}, and the Adaptive Derivative-Assembled Pseudo-Trotter ansatz 
VQE (ADAPT-VQE) models \cite{yordanov,burton,long,sun24,ramoa}.
The ADAPT-VQE method can be categorized into two classes based on the operator pool: those based on fermionic excitation operators (such as fermionic-ADAPT-VQE, fermionic excitation based (FEB) ADAPT-VQE, etc.) and those based on qubit excitation operators (such as qubit-ADAPT-VQE, qubit excitation based (QEB) ADAPT-VQE, 
coupled exchange operators (CEO) ADAPT-VQE, etc). These models adaptively add operators
during the computation process, which reduces the numbers of CNOT gates by about a factor of 10, i.e.,  from thousands to  about hundreds.   
Notably, the operator pool of CEO-ADAPT-VQE method consists of the 
linear combinations of qubit excitations, resulting in the most resource-efficient model among current ADAPT-VQE implementations. 
By incorporating the conservation of particle number into the quantum circuit design, 
our MR-UCCSD approach utilizes even smaller number of CNOT gates for  obtaining the MR wavefunction,
and the same quantum circuits are applied in the subsequent MR-UCCSD calculation, 
avoiding the need for additional quantum gates, thus 
conserving  quantum computing resources.  As listed in Table \ref{ncnot}, the present MR-UCCSD model requires the least computer resources to achieve the critical numerical accuracy of the binding energies of three molecules among all available sophisticated 
quantum computational models.

\section{summary}\label{sum}

In summary, we have developed a new efficient algorithm for MR-UCC approach,  
which is applicable 
to calculate the ground state energies for LiH, H$_6$, and BeH$_2$ in high accuracy without demanding heavy computer resources. 
The PNC circuit is applied continuously in the quantum circuit in order to  include a large number of determinants into MR wavefunctions, 
  which make possible to   induce enough higher-order correlations in 
the UCC calculations  utilizing only S and D multiplets.  
In constructing the quantum circuit,  the PNC circuit, which  generates the MR wavefunction, is again used to the
subsequent UCC calculations   utilizing the  integrating quantum computing technique with the  BCH expansion formula.   The presently proposed  algorithm  reduces significantly the number of CNOT gates 
to achieve high  accuracy of the ground state energy calculations of  LiH, H$_6$, and BeH$_2$ molecules. 
Furthermore,  the model permits  to  use the single quantum circuit along the whole bond length, including the bond-breaking region,  to achieve better accuracy than the 
chemical accuracy.
These advantages,    the numerical accuracy and the resource efficiency,    promise an applicability of our MR-UCC approach for 
the wide region of quantum computings in the NISQ era.

\begin{acknowledgements}
This work is supported by the National Natural Science Foundation of China under
Grants No. 11575120, and No. 11822504.
This work is also supported by JSPS 
KAKENHI  Grant Number  JP19K03858.  
\end{acknowledgements}

\end{document}